\documentstyle[12pt]{article}

\newcommand{\Abar}{\not{\!{\!A}}}
\newcommand{\Sbar}{\not{\!{\!S}}}
\newcommand{\pabar}{\not{\!\partial}}

\newcommand{\tr}{\mbox{tr}}
\newcommand{\Tr}{\mbox{Tr}}

\newcommand{\Dbar}{\not{\!{\!D}}}
\newcommand{\qbar}{\not{\!q}}

\def\IN{\relax{\rm I\kern-.18em N}}
\def\IR{\relax{\rm I\kern-.18em R}}
\font\cmss=cmss12 \font\cmsss=cmss12 at 7pt
\def\IZ{\relax\ifmmode\mathchoice
{\hbox{\cmss Z\kern-.4em Z}}{\hbox{\cmss Z\kern-.4em Z}}
{\lower.9pt\hbox{\cmsss Z\kern-.4em Z}}
{\lower1.2pt\hbox{\cmsss Z\kern-.4em Z}}\else{\cmss Z\kern-.4em Z}\fi}

\def\inbar{\,\vrule height1.5ex width.4pt depth0pt}
\def\IC{\relax\hbox{$\inbar\kern-.3em{\rm C}$}}

\begin{document}
\title{Some aspects of the Standard Model in
 gravitational backgrounds with torsion}
\author{Antonio Dobado \\
Departamento de F\'{\i}sica Te\'orica \\
Universidad Complutense de Madrid\\
 28040 Madrid, Spain\\ and \\
Antonio L. Maroto\\
Astronomy Centre \\
University of Sussex\\
Brighton BN1 9QH, U.K.}
\date{\today}
\maketitle
\begin{abstract}
Torsion appears in a natural way in modern formulations
of the gravitational theories. In this work we study several
aspects of the interplay between the Standard Model and a classical
gravitational background with torsion. In particular we consider the 
problem of the gauge and gravitational anomalies, $B$ and $L$ anomalies,
the effective action for the torsion and the propagation of electromagnetic
radiation in the presence of torsion.
\end{abstract}
\vskip 1.5cm
\newpage

\section{Introduction}

As  is well known, all of our positive knowledge 
about natural forces can be summarized in the Standard Model (SM) and General
 Relativity.
The Standard Model is a $SU(3)_c\times SU(2)_L \times U(1)_Y$ quantum gauge
theory which successfully describes the strong and electroweak interactions, 
even at
the high level of precision reached at LEP. The current versions and 
generalizations
of General Relativity can also in some sense be considered as gauge theories 
of the Lorentz group $SO(1,3)$. However a consistent and generally
accepted formulation of the quantum theory of gravitation is still lacking.
Thus so far our description of the gravitational phenomena is classical. 

On the other hand, one important point concerning the modern 
gravitational theories, such as supergravity \cite{sugra} or superstrings
\cite{string}
is that they consider the metric (or the vierbein) and the affine connection 
as 
different structures. This is much more natural from the mathematical 
point
of view and  leads to the appearance of torsion. Therefore, while 
waiting for a 
completely consistent theory of gravitation, it seems to be interesting 
to study 
the new effects that could appear when the Standard Model is formulated in a 
classical gravitational background  with torsion. We understand that any 
future theory should contain this approach as some kind of low-energy 
limit valid
in a regime of weak gravitational fields. In this work we will discuss 
some important
aspects of this formulation and we will make some remarks on possible 
new observable 
effects. The work is organized as follows. In Section 2 we do a brief review 
of the classical formulation of the Standard Model in  the 
presence of gravitational fields
with torsion. In order to define a proper quantum theory one important 
requirement
is the absence of gravitational and gauge anomalies. 
Anomalies arise when some symmetry of the classical lagrangian is 
spoiled by the regularization procedure in the corresponding quantum
theory, 
we deal with this issue  in 
Sections 3 and 4. In Section 5 we reconsider the Standard Model hypercharge
assignments and in Section 6 we study the effects of torsion 
on the leptonic and 
baryonic charges. In section 7 we give the effective action obtained for
the torsion and electromagnetic fields when matter is integrated out and 
study its main properties. It is then shown that quantum effects give
rise to an interaction term between these two fields which would produce 
observable 
effects. Finally in Section 8 we end with the conclusions.

\section{The Standard Model lagrangian in presence of curvature and torsion }

Let us start by introducing the  notion of torsion from a geometrical
point of view. 
Consider a pseudo-Riemannian 
space-time manifold with metric tensor $g_{\mu\nu}$. As usual, in order to
define the parallel transport of vectors, we should introduce a new object, an
affine connection, whose components we denote by 
$\hat \Gamma^{\lambda}_{\;\;\mu\nu}$. 
Such  arbitrary connection
is  in principle independent of the metric. However if we want the
lengths and angles of vectors to be invariant under parallel transport, it is
needed that the connection is metric, that is:
\begin{eqnarray}
(\hat\nabla_\lambda g)_{\mu\nu}=
\partial_\lambda g_{\mu\nu}-\hat\Gamma^{\kappa}_{\lambda\mu}g_{\kappa\nu}-
\hat\Gamma^{\kappa}_{\lambda\nu}g_{\kappa\mu}=0
\end{eqnarray}
where $\hat\nabla$ is the corresponding covariant derivative. 
This condition allows us to find the following general form for this kind 
of connections:
\begin{eqnarray}
\hat \Gamma^{\lambda}_{\;\;\mu\nu}=\Gamma^{\lambda}_{\;\;\mu\nu}
+\frac{1}{2}\left(T_{\nu\;\;\mu}^{\;\lambda}+T_{\mu\;\;\nu}^{\;\lambda}+
T_{\;\;\mu\nu}^{\lambda}\right)
\label{torchr}
\end{eqnarray}
The antisymmetric part, $T_{\;\;\mu\nu}^{\lambda}=\hat\Gamma_{\;\;\mu\nu}^{\lambda}
-\hat\Gamma_{\;\;\nu\mu}^{\lambda}$
is known as the torsion tensor and $\Gamma^{\lambda}_{\;\;\mu\nu}$
are the usual Christoffel symbols that can be obtained from the metric.
Thus we see that only for the metric and torsion free connection, both
objects (metric and connection) are not independent. 

The formulation of the SM  on curved spaces with torsion,  
can be obtained as usual 
by means of the Strong  Equivalence Principle (SEP). This principle 
 states that 
in the free falling reference frame, where the gravitational interaction 
is switched off, the physical
 laws are locally the same as in absence of 
gravitational fields. The SEP is related to the minimality of the SM
 lagrangian coupled to gravity and  yields a simple procedure to 
 couple gravity to
 any field theory built in a flat space-time.  In the following,
 we will first apply this principle to work out the gravitational
 interaction of Dirac spinors. At the end of this section 
we will also obtain the lagrangians
for scalar and gauge fields interacting with gravity.

In order to work in the path integral formalism in the following sections, 
we will be interested in the
euclidean  lagrangian 
\begin{eqnarray}
{\cal L}_M=\frac{1}{2}\left(\overline\psi \gamma^m\partial_m \psi
-\partial_{m}\overline\psi \gamma^{m}
\psi\right)
\label{dirac}
\end{eqnarray}
Before coupling gravity to this lagrangian let us introduce some notation. 
We will use 
latin 
indices $m,n...$ for 
objects referred to the locally inertial reference frame and Greek indices 
$\mu,\nu...$ for any other. If $\{ \xi^m \}$ are the coordinates in the 
privileged system and $\{ x^{\mu}\}$ the coordinates in any other then
as usual:
\begin{eqnarray}
g^{\mu \nu}(x)=e_m^{\mu}(x)e_n^{\nu}(x)\eta^{mn}
\end{eqnarray}
where $\eta^{mn}=(-,-,-,-)$ is the euclidean flat metric and 
$e_m^{\mu}(x)=\partial x^{\mu}/\partial
{\xi^m}$ is the vierbein.
In flat space-time, Dirac spinors change in the following way under
Lorentz transformations: \index{transformations!Lorentz}
\begin{eqnarray}
\psi(x) &\rightarrow & U\psi(x)=e^{\frac{i}{2}\epsilon^{mn}
\Sigma_{mn}}\psi(x) \nonumber \\
\overline\psi(x) &\rightarrow &\overline\psi(x)U^{\dagger}
=\overline\psi(x)e^{-\frac{i}{2}\epsilon^{mn}\Sigma_{mn}}
\label{lortr}
\end{eqnarray}
where $\Sigma_{mn}=\frac{i}{4}[\gamma_m,\gamma_n]$ are the hermitian 
generators of the $SO(4)$ group in the spinor representation. Notice that, 
in euclidean 
space $\psi$ and $\overline\psi$ are independent variables and the 
transformation rule of 
$\overline\psi$ is taken in such a way that $\overline\psi \psi$ is invariant. 
Therefore, the flat space-time Dirac lagrangian 
is invariant under those global transformations.

The SEP  requires the invariance of the Dirac 
lagrangian under Lorentz transformations to be not only global but also local.
With that purpose, let us introduce  a covariant derivative 
$\partial_m\rightarrow\nabla_m$ and change from the locally intertial
frame to  arbitrary coordiantes by means of the vierbien
$\nabla_\mu=e_\mu^m\nabla_m$  so that we can 
write the gauged hermitian Dirac lagrangian in the following way:
\begin{eqnarray}
{\cal L}_M=\frac {\sqrt{g}}{2} \left(\overline\psi \gamma^{\mu}\nabla_{\mu}\psi-\nabla_{\mu}
\overline\psi\gamma^{\mu}\psi\right)
\label{8dila}
\end{eqnarray}
where we have defined the Dirac matrices in curved space-time 
$\gamma^{\mu}(x)=e^{\mu}_m(x) \gamma^m$. These matrices satisfy:
 $\{\gamma^{\mu}(x),\gamma^{\nu}(x)\}=-2g^{\mu\nu}(x)$. 
The covariant derivative is defined as usual by
\begin{eqnarray}
\nabla_{\mu}=\partial_{\mu}+\Omega_{\mu}
\end{eqnarray}
where $\Omega_{\mu}$ is known as the spin connection. In order to keep the
invariance under local Lorentz transformations, $\Omega_{\mu}$ should 
transform as follows:
\begin{eqnarray}
\Omega_{\mu}\rightarrow \Omega_{\mu}'=U(x)\Omega_{\mu}U^{-1}(x)-
(\partial_{\mu}U)U^{-1}(x)
\label{spintr}
\end{eqnarray}

It can be shown that the connection components in the 
privileged reference frame  $\hat\Gamma^{a\;b}_{\;\mu}=\eta^{bc}
e^a_{\nu}(\partial_\mu e_c^{\nu}+e_c^{\lambda}
\hat \Gamma^{\nu}_{\;\;\mu\lambda}$) do 
have precisely the above transformation rule.

Following with the previous discussion, notice that 
$\{\hat\Gamma^{a\;b}_{\mu}\}$ does not 
have to be a Levi-Civita connection 
(that is, torsion free and metric), which we will denote 
$\{\Gamma^{a\;b}_{\;\mu}\}$ and therefore torsion appears
automatically. In fact, using the decomposition of the metric connection in 
eq.\ref{torchr}, we can write the 
Dirac
lagrangian in eq.\ref{8dila} in terms of the  Levi-Civita connection plus an 
additional term depending on the torsion
\begin{eqnarray}
{\cal L}_M=\sqrt{g}\bar\psi\gamma^{\mu}\left(\partial_{\mu}-\frac{i}{2}
\Gamma^{a\;b}_{\mu}\Sigma_{ab}-\frac{1}{8}S_{\mu}\gamma_5\right)\psi
\end{eqnarray}
where $S_{\rho}=\epsilon_{\mu\nu\lambda\rho}T^{\mu\nu\lambda}$. 
In conclusion, 
the lagrangian for Dirac fermions in a
curved space-time with torsion is that of a fermion in a curved 
space-time without
torsion plus a coupling between $S_\mu$ and the axial current.
 
Following
 the above arguments we can now write the corresponding expression 
for the SM matter sector:
\begin{eqnarray}
{\cal L}_M=\sqrt{g} \left(\overline {\cal Q}\Dbar^Q {\cal Q}+
\overline {\cal L}\Dbar^L {\cal L}\right)
\label{materia}
\end{eqnarray}
where:
\begin{eqnarray}
\Dbar^Q & = &\gamma^\mu D^Q_\mu= \gamma^{\mu}(\partial_{\mu}+\Omega^{Q}_{\mu}+G_\mu+W^{Q}_
{\mu}P_L+ig'B^Q_{\mu}\left(Y_L^QP_L+Y_R^QP_R\right)
+S^{Q}_{\mu}\gamma_5) \nonumber \\
\Dbar^L & = & \gamma^\mu D^L_\mu=\gamma^{\mu}(\partial_{\mu}+\Omega^{L}_{\mu}+W^{L}_{\mu}P_L+
ig'B^{L}_{\mu}(Y_L^L P_L+Y_R^L P_R)+S^{L}_{\mu}\gamma_5)
\label{dos}
\end{eqnarray}

Here we have used  the following  notation: 
$G_{\mu}$, $W_{\mu}$ and $B_{\mu}$ are the gauge fields corresponding to
$SU(3)$, $SU(2)_L$ and $U(1)_Y$ groups respectively. Quarks and lepton 
are organized in doublets 
${\cal Q}^t=({\cal U},{\cal D})$, ${\cal L}^t=({\cal N},{\cal E})$ and
 $Y^{Q,L}_{L,R}$ denote the hypercharge matrices \cite{libro}.  

As in the flat space-time case, these operators are not hermitian
due to the chiral couplings of $SU(2)_L$ and hypercharge fields. 
Thus the adjoint operators are
\begin{eqnarray}
(\Dbar^Q)^\dagger & = &\gamma^\mu \overline D^Q_\mu= 
\gamma^{\mu}(\partial_{\mu}+\Omega^{Q}_{\mu}+G_\mu+W^{Q}_
{\mu}P_R+ig'B^Q_{\mu}\left(Y_L^QP_R+Y_R^QP_L\right)
+S^{Q}_{\mu}\gamma_5) \nonumber \\
(\Dbar^L)^\dagger & = & \gamma^\mu \overline D^L_\mu=
\gamma^{\mu}(\partial_{\mu}+\overline\Omega^{L}_{\mu}+W^{L}_{\mu}P_R+
ig'B^{L}_{\mu}(Y_L^L P_R+Y_R^L P_L)+\overline S^{L}_{\mu}\gamma_5)
\label{dosad}
\end{eqnarray}

Notice that, since there is no right neutrino, the
spin connection and torsion couplings can be written as follows for 
leptonic operators:
\begin{eqnarray}
\Omega_{\mu}^L=-\frac{i}{2}\Gamma^{a\;b}_{\mu}\left(
\begin{array}{cc}
P_L\Sigma_{ab} & \; \\
\; & \Sigma_{ab}
\end{array} 
\right) 
,\; S_{\mu}^L\gamma_5=-\frac{1}{8}S_{\mu}\left(
\begin{array}{cc}
P_L\gamma_5 & \; \\
\; & \gamma_5
\end{array} 
\right)
\label{OM}
\end{eqnarray} 
where the matrices act on the flavor space. The expressions for 
$\overline \Omega$ and $\overline S$ are obtained from eq.\ref{OM} 
just replacing $P_L\rightarrow P_R$. 
 For quark operators the spin connection and torsion terms
are the same as for leptons but without the $P_{L,R}$ projectors.

Finally we will give  the lagrangians in curved space-time for the rest of
fields present in the SM.

 As the scalar fields do not
 change under Lorentz transformations, their covariant derivative
 is just an ordinary derivative. Then, according to the prescription
based on the SEP, we simply have to use the vierbein
 to perform  an arbitrary coordinate transformation. Thus, the action of 
the minimal SM symmetry breaking sector in curved space-time reads:
\begin{eqnarray}
S_{SBS}=\int d^4x \sqrt{g} \left( g^{\mu \nu}(D_{\mu}\phi)^{\dagger}
(D_{\nu}\phi)
-V(\phi)+{\cal L}_{YK}\right)
\label{min}
\end{eqnarray}
where $D_\mu=\partial_\mu+i\frac{g'}{2}B_\mu+W_\mu$ and ${\cal L}_{YK}$
is the usual Yukawa lagrangian that is not modified by the gravitational
coupling.

The Yang-Mills lagrangian in flat space-time is given by
\begin{eqnarray}
{\cal L}_{YM}=-\frac{1}{4}F^a_{mn}F_a^{mn}
\end{eqnarray}
where $a$ is a group index. 
We consider the strength tensor $F_{mn}^a$ as defined in a locally inertial 
coordinate system. $F_{mn}^a$ is a Lorentz tensor and $F^a_{mn}F_a^{mn}$ is
invariant under global and local Lorentz transformations. Therefore we only
have to transform it to an arbitrary coordinate system using the vierbein
\begin{eqnarray}
F^a_{\mu\nu}=e^m_{\; \mu}e^n_{\; \nu}F^a_{mn}
\end{eqnarray}
Thus the action for the SM gauge sector reads
\begin{eqnarray}
S_{YM}=\int d^4x \sqrt{g}\left(\frac{1}{2g_s^2}\tr G_{\mu\nu}G^{\mu\nu}+
\frac{1}{2g^2}\tr W_{\mu\nu}W^{\mu\nu}-\frac{1}{4}\tr B_{\mu\nu}B^{\mu\nu}
\right)
\end{eqnarray}

\section{The quantum Standard Model}

Up to now we have considered the classical theory. The quantization of the 
SM in curved space-time with torsion can give rise to new interesting 
effects. 
In particular, some of the above minimal 
lagrangians are not renormalizable. In fact, the one-loop calculations require
counterterms which were not present in 
the original 
lagrangian. For instance, for the 
scalar sector one needs to introduce the  counterterm $R\phi^2$ 
where $R$ is the 
scalar curvature. In addition, one 
should include in the pure gravitational
sector some counterterms that absorb the vacuum divergences, which cannot be 
discarded by the  procedures used in flat
space-time (such as normal ordering). However, the total number of new 
counterterms
that we have to add to render the theory renormalizable is finite. 
Furthermore, since symmetry is our only guiding principle
 in constructing the SM lagrangian in curved space-time,
any other non-minimal term could be included, provided it respects the 
symmetries of the theory.
All such terms are different from the minimal ones in the sense that they
 violate the SEP. The reason is that a term like $R\phi^2$
 vanishes in flat space-time, but that is not the case in a free-falling 
reference frame due to the presence of the scalar curvature. 
In contrast, the minimal couplings 
 are the same either in a flat space-time or in a
 free falling frame. 

The violation of the SEP does not 
mean a breaking of Lorentz invariance 
(provided the non-minimal terms are Lorentz scalars). 
Nevertheless, 
we will see that the anomaly effects may also violate Lorentz 
invariance, although
for consistency we will require its conservation. 
According to this discussion we conclude that the SEP is only a 
low-energy effect which will 
not be 
satisfied when higher order corrections are included in the effective 
lagrangian.

 Concerning the quantum SM there is another issue that must be taken 
into account
 in the canonical quantization procedure. In an arbitrary curved manifold,  
Poincar\'e invariance is no longer a symmetry and  $\partial/\partial t$, 
in general, is not 
a Killing field. The existence of such a Killing vector provides a 
natural definition of 
positive energy modes and therefore of creation and annihilation operators. 
As far as the vacuum is defined using  annihilation operators,  in curved 
space-time the vacuum is not unique. In this 
sense a given state which for certain observer is empty,  may 
have some particles for a 
different (accelerated) observer. These processes of particle creation are 
typical of 
Quantum Field Theory (QFT) in curved space-time
and they have been extensively studied \cite{birrell}, \cite{parprod}.

 In addition the definition of an 
$S$-matrix requires a time parameter with respect to which 
we can define asymptotically free states in the remote past and future. In 
fact, in Minkowski space-time, particles can be well separated 
before and after the interaction. However, in curved space-time 
this situation does not take place in general and, as a consequence,
it is not always possible to define an $S$-matrix.

Finally it is well known that  chiral theories (like the Standard Model)
in four dimensions are potentially inconsistent due to gauge and mixed
gauge-gravitational anomalies. In flat space-time, the assignment of
hypercharges for the different fermions is done is such a way that the 
contributions to the anomalies of some fermion field is exactly cancelled
by the contributions of the rest of fermions. However when gravity is 
introduced things can change since there is new contributions to the 
anomalies coming from curvature and torsion.
In the next sections we will study 
these new terms 
and we will extract the  conditions needed for their cancellation.

\section{The Standard Model anomalies}

Let us consider the effective action for the gauge fields, the vierbein
and the spin-connection that is obtained after the functional integration
of the fermionic fields in the SM:
\begin{eqnarray}
e^{-W[A,\Omega,e]}=\int [d\psi  d\overline\psi]
e^{-\int d^4x{\cal L}_{SM}}
\label{ea}
\end{eqnarray}
Here $A$ denotes the gauge fields and $\psi$ all the fermions.
This effective action contains all the information about the quantum
effects of the matter fields on the gravitation and gauge interactions.
In particular, from eq.\ref{ea} it is possible to extract, 
particle creation rates, gauge and gravitational anomalies,
 new interaction
terms, etc.

In order to calculate the SM anomalies we will follow 
the  standard Fujikawa method \cite{fujikawa}. There exist
a gauge anomaly whenever the EA is not gauge invariant. 
Let us consider as an example the  $SU(N_c)$ gauge transformations. 
The 
effective action, being a functional in the gauge fields, 
 may have an anomalous variation given by
\begin{eqnarray}
\delta_\theta W
=-\int d^4x \sqrt{g}i\theta^b(D_{\mu}\langle j^{\mu}_c \rangle)^b=
-\int d^4x \sqrt{g}i\theta^a A^a(x)
\label{caea}
\end{eqnarray}
where $D_{\mu}=\nabla_{\mu}+[G_\mu,\cdot]$, $\theta$ is the transformation
parameter and $j^{a\mu}=\overline{\cal Q}\gamma^\mu\Lambda^a{\cal Q}$
the color gauge current.  
This transformation comes from the change in the integration measure
in eq.\ref{ea}, 
since the classical action is gauge invariant.

The anomaly $A^a(x)$ in the above equation is in general a 
divergent object. The corresponding renormalized expression is obtained 
 by using the regularizing operators defined by: 
\begin{eqnarray}
H_{\psi}^{{\cal Q},{\cal L}}&=&
(i\Dbar^{{\cal Q},{\cal L}})^\dagger(i\Dbar^{{\cal Q}, {\cal
L}})= {\cal D}_{\mu}^{{\cal Q},{\cal L}}{\cal D}^{\mu}_{{\cal Q},{\cal L}}
+X^{{\cal Q},{\cal L}}  \nonumber \\
H_{\overline\psi}^{{\cal Q},{\cal L}}&=&
(i{\Dbar}^{{\cal Q},{\cal L}})(i{\Dbar}^{{\cal Q},
{\cal L}})^\dagger =\overline {\cal D}_{\mu}^{{\cal Q},{\cal L}} 
\overline {\cal D}^{\mu}_{{\cal Q},{\cal L}}+\overline X^{{\cal Q},{\cal L}}
\label{8regop}
\end{eqnarray}
with $X=\gamma_5 S^{\mu}_{\; ;\mu}+2
S_{\mu}  S^{\mu}-
\frac{1}{4}[\gamma^{\mu},\gamma^{\nu}][d_{\mu},d_{\nu}]$
and ${\cal D}_{\mu}=D_{\mu}
-\frac{1}{2}\gamma_5[\gamma_{\mu},\gamma^{\nu}]S_{\nu}$
whose eigenvalues are gauge and local Lorentz invariant. As before, a bar
means that left and right projectors have to be exchanged. 

The usual renormalization prescription for the anomaly in the
Fujikawa method is given by \cite{doma}:
\begin{eqnarray}
A^a_{ren}(x)=\frac{1}{(4\pi)^2}\Tr(\Lambda^a (a_2(H_\psi^{\cal Q},x)
-a_2(H_{\bar\psi}^{\cal Q},x)))
\label{aren}
\end{eqnarray}

The second coefficient in the heat-kernel expansion in curved space-time for 
the operators in eq.\ref{8regop},
which  is the only relevant for the anomaly
calculation, reads 
\begin{eqnarray}
a_2(H_{\psi},x)
&=&\frac{1}{12}[{\cal D}_{\mu},{\cal D}_{\nu}]
[{\cal D}^{\mu},{\cal D}^{\nu}]
+\frac{1}{6}[{\cal D}_{\mu},
[{\cal D}^{\mu},X]]+
\frac{1}{2}X^2-
\frac{1}{6}RX \nonumber \\
&-&\frac{1}{30}R_{;\mu}^{\;\;\mu}+\frac{1}{72}R^2
+\frac{1}{180}(R_{\mu \nu \rho
\sigma}  R^{\mu \nu \rho \sigma} -R_{\mu \nu} R^{\mu \nu})
\label{a2p}
\end{eqnarray}
and
$a_2(H_{\overline\psi},x)$ has similar expression but with the barred
quantities. 

From the previous discussion, we obtain the anomalous Ward identity
\begin{eqnarray}
(D_{\mu}\langle j^{\mu} \rangle)^a=A^a_{ren}
\end{eqnarray}

Thus we see that the presence of the anomaly represents a failure in
the conservation of the gauge currents.
Such non-conservation  would destroy the consistency
 of the model and  then it is necessary that the new terms 
depending on the curvature and torsion cancel. This could impose new 
constraints to 
the SM hypercharges. 
Up to now  we have only considered  the anomalous Ward
identities in the $SU(N_c)$ case, the $SU(2)_L$ and $U(1)_Y$ results 
are obtained in a similar
way. 

The final explicit results for the different anomalous Ward identities are
obtained just by taking the traces in Lorentz and internal 
indices in eq.\ref{aren}. Thus for $SU(N_c)$:
\begin{eqnarray}
(D_{\mu}\langle j^{\mu}_c\rangle)^a=
-\frac{1}{32\pi^2}g_sg'\epsilon^{\mu \nu \rho \sigma}G^a_{\mu \nu}
B_{\rho\sigma}\sum_{u,d}(y_L-y_R)
\label{color}
\end{eqnarray}
The $SU(2)_L$ gauge current is $j^{a\mu}_L=\overline{\cal Q}\gamma^\mu T^aP_L
{\cal Q}+({\cal Q}\rightarrow {\cal L})$ and the corresponding anomaly yields:
\begin{eqnarray}
(D_{\mu}\langle j^{\mu}_L\rangle)^a=
-\frac{1}{32\pi^2}gg'\epsilon^{\mu \nu \rho \sigma}
W^a_{\mu \nu}
B_{\rho\sigma}\left(N_c\sum_{u,d}y_L+\sum_{\nu,e}y_L\right)
\label{left}
\end{eqnarray}
The $U(1)_Y$ gauge current is $j^\mu_Y=\overline{\cal Q}\gamma^\mu(Y^Q_LP_L+
Y^Q_RP_R){\cal Q}+({\cal Q}\rightarrow {\cal L})$ and the anomaly:
\begin{eqnarray}
D_{\mu}\langle j^{\mu}_Y \rangle & =
&\frac{1}{32\pi^2}\left( -\frac{1}{24}
\epsilon^{\rho \sigma \gamma \delta} R_{\mu \nu \rho
\sigma} R^{\mu \nu}_{\; \; \;\gamma\delta}
\left[
N_c\sum_{u,d}(y_L-y_R)+\sum_{\nu,e}(y_L-y_R)\right ]
   \right.  \nonumber \\
 & + &  \frac{g_s^2}{2}\epsilon^{\mu \nu \rho \sigma}G^a_{\mu \nu}G^a_{\rho
\sigma}\sum_{u,d}(y_L-y_R)+
 \frac{g^2}{4}\epsilon^{\mu \nu \rho \sigma}W^a_{\mu
\nu}W^a_{\rho \sigma}\left[ N_c\sum_{u,d}y_L  
+\sum_{\nu,e}y_L\right] \nonumber \\ 
  &+&  \left. {g'}^ 2 \epsilon^{\mu \nu \rho \sigma}B_{\mu
\nu}B_{\rho \sigma}
\left[N_c\sum_{u,d}(y_L^3-y_R^3)+\sum_{\nu,e}(y_L^3-y_R^3)\right]\right ) 
\label{hiper}
\end{eqnarray}

Thus we find the same result as in flat space-time for the non-abelian
currents. However the abelian $U(1)_Y$ anomaly does get contributions from
the curvature and the torsion, thus imposing new conditions on the hypercharges
for the cancellation. The torsion contribution can be seen to be  nothing but
a total derivative and therefore can be removed by adding suitable
counterterms to the lagrangian, for that reason we have not written it
in the final result.

As we have mentioned above, the SEP states that any theory in curved space-time
should be invariant under local Lorentz transformations. 
We consider the 
possible
violation of this symmetry due to quantum effects when 
chiral fermions are present. As we will see, 
whenever abelian chiral gauge fields are present, as  
is the case of the hypercharge field, local Lorentz invariance is 
in principle broken.   

As we saw in Sect.2, the classical Dirac lagrangian 
in curved space-time is invariant under
the $SO(4)$  transformations given in eq.\ref{lortr} and eq.\ref{spintr}  
(for Euclidean signature).
 Therefore we can calculate
the gravitational anomalies as gauge anomalies of the Lorentz group. 
The only difference
is the appearance of an additional field, the vierbein,
which also transforms under this group.

Following the same steps as in the previous section, we can obtain the
corresponding anomalous Ward identities. 

We can write this result more conveniently using
\begin{eqnarray}
\frac{\delta W}{\delta \Gamma^{a\; b}_{\; \mu}}&=&-\frac{i}{4}\langle \bar\psi(\gamma^{\mu}\Sigma_{ab}+
\Sigma_{ab} \gamma^{\mu})\psi \rangle=-\frac{i}{2}\langle j_{ab}^{\; \; \mu} \rangle \nonumber \\
\langle j^{\mu} \rangle&=&\langle j_{ab}^{\; \; \mu} \Sigma^{ab} \rangle 
\end{eqnarray}
and the explicit form of the $SO(4)$ gauge 
covariant derivative that is nothing but:  
$D_{\mu}=\nabla_{\mu}+
[\Omega_{\mu},\cdot]$.

Following the same steps as in the gauge case  we find the 
anomalous Ward identity:
\begin{eqnarray}
A^{ab}_{ren}(x)=-(D_{\mu}\langle j^{\mu} \rangle)^{ab}(x)+i(T^{ab}(x)-T^{ba}(x))
\end{eqnarray}
where $T_{ab}=e_{b\mu}\delta W/\delta e^a_{\;\;\mu}$ is the expectation 
value of  the energy-momentum tensor in the presence of the background fields.

The renormalized anomaly  is:
\begin{eqnarray}
A^{ab}_{ren}(x)=\frac{1}{(4\pi)^2}\tr\left[\Sigma^{ab}\left(a_2(H_{\psi}^{\cal Q},x)-
a_2(H_{\overline\psi}^{\cal Q},x)\right)\right]+({\cal Q}\rightarrow {\cal L})
\end{eqnarray}

After a lengthy calculation we arrive to the final expression for the 
Lorentz anomaly, its explicit expression is:
\begin{eqnarray}
A^{mn}_{ren} & = & \frac{g'}{32\pi^2}\left(
\frac{1}{6}\epsilon^{mnab}R_{\mu \nu ab}B^{\mu \nu}
+\frac{1}{6} \left(B_{\alpha}^{\;\;n}S^{\alpha;m}
-B_{\alpha}^{\;\; m}S^{\alpha ; n}\right)
\right. \nonumber \\
 & - &
\frac{1}{24}\epsilon^{mnab}\left(B_{ac}S^cS_b+B_{ab}S^2\right)
-\frac{1}{6}\epsilon^{mnab}B_{ab}R 
-\frac{1}{2}S^{\mu}_{\; ;\mu}B^{mn}\nonumber \\
 & - & 
\frac{1}{3}\epsilon^{mnab}\Box B_{ab}
+\frac{1}{3}\left(S_{\alpha}B^{\alpha m}\right)^{;n}
-\frac{1}{3}\left(S^m B^{\mu n}-S^n B^{\mu m}\right)_{;\mu}
\nonumber \\
&-& \left.\frac{1}{3}\left(S_{\alpha}B^{\alpha n}\right)^{;m}\right)
\left(\sum_{u,d}N_c(y_L-y_R)+
\sum_{\nu,e}(y_L-y_R)\right)
\label{lal}
\end{eqnarray}
Notice that pure gravity terms do not arise, in agreement with the result 
that there are no pure 
gravitational
anomalies in four dimensions. Observe also that all the terms depend on 
the abelian  $B_{ab}$ field,
whereas there is no contribution from non-abelian gauge fields. 
Finally, the cancellation condition agrees with that of 
eq.\ref{8c3} which ensures the vanishing of the gravity terms in the $U(1)_Y$ 
anomaly. This condition is satisfied with the usual hipercharge assignment 
in the SM.

\section{Anomaly cancellation and charge quantization in the SM}

From the above computation of the SM gauge and gravitational anomalies
it is very easy to read the conditions that must be set on the hypercharges
in order to cancel those anomalies thus making possible a proper definition 
of the SM as a 
QFT. The three first conditions in eqs.\ref{color}, \ref{left} and \ref{hiper} 
are the same as those obtained in flat 
space-time.
However notice the appearance of terms depending on the curvature in the 
third condition, 
which did not occur in the case of
non-abelian gauge fields. 
The new terms that were not present in flat space-time  
impose a new cancellation 
condition, namely, the vanishing of the sum of all hypercharges
\begin{eqnarray}
N_c\sum_{u,d}(y_L-y_R)+\sum_{\nu,e}(y_L-y_R)=0
\label{8c3}
\end{eqnarray}
The conditions for the cancellation
of gauge anomalies in flat space-time, 
together with the  gauge invariance of the SM Yukawa sector,   
allows us to fix all the hypercharges up to
a normalization constant \cite{Ramond}. However, we have just seen that, 
in curved space-time,  
we have an additional constraint on the hypercharges, 
eq.\ref{8c3}, coming both from 
the curvature terms in the $U(1)_Y$ 
anomaly and from the local Lorentz anomaly. Within the minimal SM, this  condition is 
compatible with the others. 
However it is possible to take a different point of view and, 
without assuming any specific
symmetry breaking sector, try to fix the hypercharges. Then, the above
conditions  form a set of four equations for five 
unknowns $y_L^{\nu}=y_L^{e},
\;y_L^{u}=y_L^{d},\; y_R^{e},\; y_R^{u}\; ,y_R^{d}$. Let us try to solve the
system explicitly and check whether they determine all the hypercharges
up to a normalization factor. First, we note that the four equations
can be reduced to a single one in terms of one variable 
for $y^d_R \ne 0$ (if $y^d_R=0$ all the hypercharges vanish, 
which is unphysical):
\begin{eqnarray}
1+\left(\frac{y^u_R}{y^d_R}\right)^3+\frac{21}{6}\left(
\frac{y^u_R}{y^d_R}\right)^2+\frac{21}{6}\frac{y^u_R}{y^d_R}=0
 \end{eqnarray}
It is not difficult to see that 
there are three real solutions,
\begin{eqnarray}
\frac{y^u_R}{y^d_R}=-1,\; -2,\; -\frac{1}{2}
\end{eqnarray}
The rest of the hypercharges can be obtained in terms of one
of them in a strightforward way.  
Hence, there are only three possible sets of hypercharges, up to 
a global normalization factor. We have listed them in Table 8.1.  
The first solution, whose normalization is arbitrary,  
together with the usual weak isospin assignment
$Q=T_3+Y$
 implies that the right component of the electron is chargeless. 

The second 
set is the usual hypercharge assignment in the SM, 
normalizing as usual $Q_{\cal E}=-1$. The 
third solution,
keeping the same normalization,
leads to  different electric charges for the left and right components of the
quark fields and therefore to
 chiral electromagnetism. In conclusion, gauge and local Lorentz  invariance
together with some physical constraints such as vector electromagnetism, 
allows us to fix all the hypercharges up to global normalization.

\begin{table}
\begin{center}
\begin{tabular}{|c|c|c|c|c|c|c|c|c|c|} \hline\hline
$  $&${\cal U}_R$&${\cal U}_L$&${\cal D}_R$&${\cal D}_L$&${\cal N}_L$
&${\cal E}_R$&$ {\cal
E}_L$\\  \hline
$1^{st}$ set &$y$&$0$&$-y$&$0$&$0$&$0$&$0$\\ 
\hline
$2^{nd}$ set& $2/3 $    & $  1/6$   &$-1/3 $    &   $1/6$   & $-1/2$    &    $-1$   & $-1/2$  \\
\hline $3^{rd}$ set& $-1/3$&$1/6$&$2/3$&$1/6$&$-1/2$&$-1$&$-1/2$\\
\hline \end{tabular}
\end{center}
\centerline{\footnotesize {\bf Table 8.1:} Hypercharge assignments.
\index{hypercharge!assignments}}
\end{table}

\section{The leptonic and baryonic anomalies}

Following the above methods we will study in this section the anomalies 
appearing in the leptonic and baryonic currents. Those anomalies correspond
to global symmetries of the classical SM and then they do not
lead to any inconsistency in the quantum theory. Moreover they
could lead to new observable effects. In flat space-time neither the 
baryon number $B$ nor the leptonic number $L$ are conserved because of the
axial anomaly. However, those anomalies equal so that $B-L$ is still conserved
in the presence of the anomaly. 

The baryon and lepton 
number currents:
\begin{eqnarray}
j^\mu_B=\frac{1}{N_c}\overline {\cal Q} \gamma^\mu {\cal Q} \; , 
j^\mu_L=\overline {\cal L} \gamma^\mu {\cal L}
\end{eqnarray}
are classically conserved, i.e $\nabla_\mu j^\mu_{Q,L}=0$, where 
$\nabla_{\mu}$ is
the Levi-Civita covariant derivative. However these conservation laws are 
violated due to
quantum effects. The corresponding anomalies can be obtained using
Fujikawa method based on the operators defined in eq.\ref{8regop}.
 The gaussian regulators associated to these operators respect 
the gauge and 
local Lorentz symmetries of the theory. This procedure yields for 
the anomalies the 
following results (for euclidean signature):
\begin{eqnarray}
\nabla_{\mu}j^{\mu}_B=\frac{1}{32\pi^2}\epsilon^{\mu \nu \alpha
\beta}\left(\frac{g^2}{2} W^a_{\mu \nu}
W^a_{\alpha\beta}+g'^2B_{\mu\nu}B_{\alpha\beta}\sum_{u,d}(y_L^2-y_R^2)\right)
\label{ban}
\end{eqnarray}
and 
\begin{eqnarray}
\nabla_{\mu}j^{\mu}_L=\frac{1}{32\pi^2}\left \{ -\frac{\epsilon^{\alpha \beta
\gamma \delta}}{24} R_{\mu \nu \alpha \beta}
R^{\mu\nu}_{\; \; \;\gamma\delta}+\frac{\epsilon^{\alpha \beta
\gamma \delta}}{48}S_{\beta ; \gamma}S_{\delta ;
\alpha} +\epsilon^{\alpha \beta
\gamma \delta}\left(\frac{g^2}{2}W^a_{\gamma \delta}
W^a_{\alpha\beta}\right. \right. \nonumber \\
\left. \left.+g'^2B_{\gamma
\delta}B_{\alpha\beta}\sum_{\nu,e}(y_L^2-y_R^2)\right)+
\frac{1}{6} \Box S^{\alpha}_{\; ;\alpha}+ \frac{1}{96}\left(S^{\alpha}S^
{\nu}S_{\alpha}\right)_{;\nu} -\frac{1}{6}\left(R^{\nu
\alpha}S_{\alpha}-\frac{1}{2}RS^{\nu}\right)_{; \nu}\right\}
\label{lan}
\end{eqnarray}
The resulting  lepton anomaly has terms depending on the curvature and the 
torsion that appear due to the absence of one of the chirallity components 
of the neutrino
field. Such terms are not present in the baryonic anomaly since quarks have 
both chirallity
components. Thus in contrast with flat space-time, $B-L$  is spoiled in 
the presence 
of curvature and in principle also in presence of torsion. 

Therefore we have three different kinds of potential contributions to the 
lepton 
anomaly. First we
have a $\epsilon^{\alpha \beta
\gamma \delta} R_{\mu \nu \alpha \beta}
R^{\mu\nu}_{\; \; \;\gamma\delta}$ term where 
$R^{\mu\nu}_{\; \; \;\gamma\delta}$ is the 
curvature associated to the Levi-Civita spin connection. This  topological 
term could give rise to actual contributions to the $L$ violation through 
the so called 
gravitational instantons  that would be relevant in the 
context of quantum 
gravity \cite{ibanez}. Second, there is also the 
well known $SU(2)_L$ and
$U(1)_Y$ gauge contributions to the anomaly \cite{shapos}. Finally, we also 
have the possible
 torsion contribution in 
which we are interested.

However the  
torsion contribution  to the anomaly is a four-divergence and it can 
be absorbed  in the
redefinition of the  lepton current as follows:
\begin{eqnarray}
\tilde j^{\mu}_L=j^{\mu}_L
-\frac{1}{32\pi^2} \left( \frac{1}{6}S_{\alpha}^{\; ;\alpha\mu}
+\frac{1}{96}S^{\alpha}S^{\mu}S_{\alpha}
-\frac{1}{6}\left(R^{\mu\alpha}S_{\alpha}
-\frac{1}{2}RS^{\mu}\right)
+\frac{1}{48}\epsilon^{\mu\beta\gamma\delta}S_{\beta;\gamma}S_{\delta}
\right)  
\label{redcur}
\end{eqnarray} 
In the absence of the other contributions to the anomaly this 
new current is 
conserved, that is, 
$\nabla_{\mu}\tilde j^{\mu}_L=0$. Thus we observe that the possible effects 
of the torsion can be 
eliminated away by this redefiniton of the 
lepton current. Notice that the new definiton, being gauge and 
local Lorentz invariant, is physically meaningful and should 
satisfy the appropriate Ward identities for a properly defined lepton current.

An alternative way to show that torsion does not contribute to the
lepton anomaly is based on a different choice of regulator. In fact
let us take as regulators for the anomalies the operators 
$i\tilde{\Dbar}^\dagger i\tilde{\Dbar}$ 
and $i\tilde{\Dbar} i\tilde{\Dbar}^\dagger$, where $\tilde{\Dbar}$ is
the ${\Dbar}$ operator in which the torsion field has been set to
zero. These operators respect all the gauge and local Lorentz
symmetries of the theory and therefore they are also valid
as regulators. However they do not depend on torsion, this implies 
that the regulated anomalies cannot depend on torsion as expected.

\section{The EA for torsion and electromagnetism}

The effective action defined in eq.\ref{ea} is a functional in the vierbein
and the connection, i.e, in the curvature and the torsion. In addition
it also depends on the gauge field. In this section we will concentrate
in the contributions to the EA coming only from the electromagnetic and
torsion fields, from them it could be possible to obtain some
phenomenological effects as we will see. Therefore we will omit the 
curvature dependence working in a flat space-time and also ignore
the effects of the rest of the gauge fields.

In this case, we will get for the EA, considering only the electronic family
in SM:
\begin{eqnarray}
W[A,S]=S_{cl}[A,S]+i\Tr\;\log\left(i\Dbar-m-\frac{1}{8}\Sbar\gamma_5\right)
\end{eqnarray}
where $S_{cl}[A,S]$ denotes the classical action which includes the Maxwell
term for the EM field, as well as the corresponding action for torsion.  
The Dirac operator
coupled to EM is defined as usual: $D_\mu=\partial_\mu-ieA_\mu$.
We can formally expand  the logarithm to generate a series in the
interaction terms $S_\mu$ and $A_\mu$:
\begin{eqnarray}
W[A,S]=S_{cl}[A,S]+i\sum_{k=1}\frac{(-1)^k}{k}\Tr\left((i\pabar-m)^{-1}(e\Abar
-\frac{1}{8}\Sbar\gamma_5)\right)^k
\end{eqnarray}
The Dirac propagator is defined as usual:
\begin{eqnarray}
(i\pabar-m)^{-1}_{xy}=\int d\tilde qe^{-iq(x-y)}\frac{\qbar+m}{q^2-m^
2+i\epsilon}
\end{eqnarray}
where the functional traces $\Tr$ are evaluated in  dimensional
regularization with $D=4-\epsilon$ and $d\tilde q=\mu^\epsilon d^Dq/(2\pi)^D$.
The result will contain divergent local pieces together with
finite local and non-local terms. Let us first give the results for the
divergences:
\begin{eqnarray}
W_{div}[A,S]=\frac{\Delta}{(4\pi)^2}\int d^4x\left(-\frac{e^2}{3} 
F_{\mu\nu}F^{\mu\nu}
-\frac{1}{192}S_{\mu\nu}S^{\mu\nu}+\frac{m^2}{16}S_\mu S^\mu\right)
\label{div}
\end{eqnarray}
where $S_{\mu\nu}=\partial_\mu S_\nu-\partial_\nu S_\mu$,  
$\Delta=N_\epsilon+\log (\mu^2/m^2)$ with the poles parametrized as
$N_\epsilon=2/\epsilon+\log 4\pi-\gamma$ and
$\mu$ the renormalization scale. The first divergent term will be absorbed
in the redefinition of the photon field. In order to absorb the divergences
depending on torsion, it is neccessary that the classical action contains
a kinetic and a mass term for the torsion field. Thus we see that torsion
behaves like a massive abelian gauge field. Even if we had started from
a theory with non-propagating torsion (such as Einstein-Cartan theory),
we would generate a kinetic term  due to the interaction
with the fermions. 

The rest of finite contributions to the EA are in general
difficult to obtain, however we can study some of them in some
particular limits. Thus we will consider those terms with two photon fields
and one torsion field in the masless limit for fermions and for slowly
varying torsion fields, i.e, $p^2_A>>m^2$ and $p^2_A>>p^2_S$, where
$p_A$ and $p_S$ denote the momenta of photon and torsion respectively.
To the lowest order in the number of torsion derivatives, we have:
\begin{eqnarray}
W^{AAS}_0[A,S]=-\frac{e^2}{4(4\pi)^2}\int d^4x \epsilon_{\mu\nu\rho\sigma}
F^{\rho\mu}A^\sigma S^\nu
\end{eqnarray}
This term is gauge invariant but only to the lowest order, that is, for
constant torsion. It is local and finite since all the possible divergences
are those in $W_{div}$. The next order in the expansion with one
torsion derivative reads:
\begin{eqnarray}
W^{AAS}_1[A,S]=\frac{e^2}{2(4\pi)^2}\int d^4x d^4y d\tilde p\frac{e^{ip(x-y)}}
{p^2}\epsilon^{\alpha\beta\mu\nu}\partial_\mu\partial_\lambda A^\lambda(y)
A_\nu(x)\partial_\alpha S_\beta(x)
\end{eqnarray}
This term is non-local and is not gauge invariant, however if we 
add $W^{AAS}_0$ we recover gauge invariance even 
for arbitrary torsion fields. 
For constant
$S_\mu$ the term $W^{AAS}_1$ vanishes and $W^{AAS}_0$ has the precise form
of the term needed to explain \cite{anis}  the  anisotropy in cosmological
electromagnetic propagation recently claimed by Nodland and Ralston 
\cite{Nodland}. 
However two main difficulties appear in this explanation:

i) The term has been obtained assuming that $p_A^2>>m^2$, i.e. it is
only valid for highly energetic photons. However the Nodland-Ralston
effect was found in radio galaxies and therefore one expects $W^{AAS}_0$
not to be very relevant in that energy range. However the very same 
fact suggests to look for such an effect in gamma-rays sources.

ii) More important is the origin of the cosmological slowly varying
torsion field. From eq.\ref{div} we see that torsion behaves like
a massive gauge field with mass values around Planck scale. This fact would
avoid the generation of long range torsion fields. However it has been
suggested that the true vacuum of the theory might have a non vanishing
vacuum expectation value for the torsion pseudotrace, thus generating the
cosmological  background \cite{Andrianov} required in the above
explanation.

\section{Conclusions}
We have reviewed some of the consequences of formulating the SM in 
a curved space-time with torsion. In particular, we have shown that
the presence of torsion is not incompatible with the cancellation of
gauge and gravitational anomalies and thus, the SM can be consistently
formulated in a gravitational background with torsion. In addition, 
although the leptonic anomaly is not
modified by the presence of torsion, it gets new contributions from 
the Levi-Civita curvature and as a consequence $B-L$ is no longer 
conserved
in the presence of gravity.
Finally we have shown that quantum effects generate a kinetic and a mass
term for the torsion fields of the same form as those of
 massive abelian gauge fields. 
In addition, when we have torsion together with an electromagnetic 
background, the quantum effects of matter fields induce a new
coupling of EM with torsion that in principle could have 
phenomenological
consequences.

{\bf Aknowledgments:}
We thank L. Alvarez-Gaum\'e, A.A. Andrianov and I.L. Shapiro for
useful discussions. 
This work has been partially supported by the Ministerio de Educaci\'on y
Ciencia (Spain)(CICYT AEN96-1634). A.L.M is supported 
by SEUID-Royal Society.

\end{document}